\documentclass[twocolumn]{jpsj2}

\usepackage{graphicx}

\title{Conserving Gapless Mean-Field Theory for Bose-Einstein Condensates}

\author{Takafumi Kita}
\inst{Division of Physics, Hokkaido University, Sapporo 060-0810, Japan}

\recdate{\today}

\abst{%
We formulate a conserving gapless mean-field theory
for Bose-Einstein condensates on the basis of a Luttinger-Ward thermodynamic functional.
It is applied to a weakly interacting uniform gas 
with density $n$ and $s$-wave scattering length $a$
to clarify its fundamental thermodynamic properties.
It is found that the condensation here occurs as a first-order transition.
The shift of the transition temperature $\Delta T_c$
from the ideal-gas result $T_{0}$ is positive and given to the leading order by
$\Delta T_c \!=\! 2.33a n^{1/3}T_0$,
in agreement with a couple of previous estimates.
The theory is expected to form a new theoretical basis for
trapped Bose-Einstein condensates at finite temperatures.
}

\kword{%
Bose-Einstein condensation, mean-field theory, gapless excitation, conservation laws}

\begin{document}
\sloppy
\maketitle

Much effort has been devoted recently to calculating
the leading-order correction $\Delta T_c\!\equiv\! T_c\!-\! T_0$
to the transition temperature of a weakly interacting
uniform Bose gas;
see refs.\ \citen{Arnold01} and \citen{Yukalov04} for an overview.
However, the results differ substantially among different approaches,
with $\Delta T_c$ ranging from negative to positive values.
Unlike the quantities at $T\!=\! 0$, 
perturbational calculations are difficult for $\Delta T_c$.
This situation also highlights the absence
of an established 
mean-field theory for Bose-Einstein condensates (BEC),
corresponding to the Bardeen-Cooper-Schirieffer theory of superconductivity,
with which one could estimate $\Delta T_c$ easily.
Thus, we still do not have a quantitative description
of BEC at finite temperatures, particularly near $T_c$.

As was reported by Hohenberg and Martin in 1965 \cite{HM65}
and emphasized recently by Griffin \cite{Griffin96},
there are a couple of key words that can be used 
in formulating a mean-field theory for BEC:
``conserving'' and ``gapless.''
Here, application of the conventional Wick-factorization procedure 
to the Hamiltonian of interacting Bose particles,
known as the Hartree-Fock-Bogoliubov (HFB) 
theory \cite{Griffin96,GA59},
fails to satisfy the Hugenholtz-Pines theorem \cite{HP59}
giving rise to an
unphysical energy gap in the excitation spectrum. 
Thus, it has become customary to introduce a further approximation,
now called the ``Popov'' approximation \cite{Griffin96}, 
of completely ignoring the anomalous quasiparticle pair correlation
to recover a gapless excitation.
However, when the same approximation is adopted for dynamics, it
does not satisfy various conservation laws
as required.
Also, it is not clear whether it is permissible
to neglect the pair correlation,
which has played an essential
role in the pioneering perturbation theory of Bogoliubov
\cite{Bogoliubov47}; see also refs.\ \citen{LHY57} and \citen{Fetter72}.
Recently, Proukakis {\em et al}.\ \cite{Proukakis98}
have presented an improved gapless theory
with the pair correlation. However,
it still does not satisfy the conservation laws.

On the basis of these observations, we here formulate a new mean-field theory
for BEC with the desired conserving gapless character.
As first shown by Baym\cite{Baym62} and
discussed in detail by Hohenberg and Martin,\cite{HM65}
the conservation laws are well satisfied in ``$\Phi$ derivative approximations''
where the self-energy $\hat{\Sigma}$ is given as a derivative of some functional
$\Phi\!=\!\Phi[\hat{G}]$ with respect to Green's function $\hat{G}$.
This relationship between $\hat{\Sigma}$ and
$\hat{G}$ holds exactly 
for Fermi systems, as shown by Luttinger and Ward.\cite{LW60}
Indeed, $\Phi$ was first introduced by Luttinger and Ward as a part of the
exact thermodynamic functional $\Omega\!=\!\Omega[\hat{G}]$. It was
used later by Baym\cite{Baym62} to give the criterion for
obtaining dynamical equations
with conservation laws.
We hence ask the question here: 
Can we construct a Luttinger-Ward thermodynamic functional for BEC
within a mean-field approximation
that also satisfies the
Hugenholtz-Pines theorem?
This is indeed possible as will be shown below.
The predictions of the resulting mean-field theory for the uniform system
will be presented later with an expression for $\Delta T_c$.
We put $\hbar\!=\! k_{\rm B}\!=\! 1$ throughout.

As usual, we express the field operator $\psi({\mib r})$
as a sum of the condensate wave function $\Psi({\mib r})$
and the quasiparticle field $\phi({\mib r})$.
We also adopt Beliaev's Green's function approach \cite{Beliaev58}
and define our Matsubara 
Green's function in Nambu space as
\begin{eqnarray}
\hat{G}({\mib r},{\mib r}',\tau)\equiv -  \hat{\tau}_{3} \,\langle T_{\tau}
\mbox{\boldmath $\phi$}({\mib r},\tau)
\mbox{\boldmath $\phi$}^{\dagger}({\mib r}')\rangle\, ,
\label{hatG}
\end{eqnarray}
where $\hat{\tau}_{3}$ is the third Pauli matrix and
\begin{eqnarray}
\mbox{\boldmath $\phi$}=\left[
\begin{array}{c}
\phi \\ \phi^{\dagger}
\end{array}
\right]  ,
\hspace{5mm}
\mbox{\boldmath $\phi$}^{\dagger}=
[\, \phi^{\dagger} \,\, \phi\,] \, .
\label{phi}
\end{eqnarray}
The factor $\hat{\tau}_{3}$ in eq.\ (\ref{hatG})
is usually absent \cite{HM65,Griffin96} but
essential for the present purpose.
Indeed, the eigenvalue problem
for $\hat{G}^{-1}$ then becomes equivalent to the 
Bogoliubov-de Gennes equation for quasiparticles,
as seen below in eqs.\ (\ref{Dyson}) and (\ref{BdG}),
so that the logarithmic term in eq.\ (\ref{LWF}) can be written entirely
with respect to the quasiparticles.

Using the Fourier transform $\hat{G}({\mib r},{\mib r}',\omega_{n})$ with
$\omega_{n}$ being the Matsubara frequency,
we write down our Luttinger-Ward functional
$\Omega\!=\!\Omega (\Psi,\Psi^{*},\hat{G})$ as
\begin{eqnarray}
&&\hspace{-8mm}
\Omega = 
\int \! \Psi^{*}({\mib r}) (H_{0}\!-\!\mu)\Psi({\mib r}) 
\, d{\mib r}
\nonumber \\
&& \hspace{-1mm}
+ \frac{1}{2\beta} \sum_{n=-\infty}^{\infty} {\rm Tr}\left[
\ln (\hat{H}_{0}\!+\!\hat{\Sigma} \!-\! \mu \hat{\tau}_{3} 
\!-\! i\omega_{n}\hat{1})
+\hat{G}\hat{\Sigma}\right]\hat{1}(\omega_{n})
\nonumber \\
&& \hspace{-1mm}
+ \, \Phi
\, .
\label{LWF}
\end{eqnarray}
Here,
$H_{0}$ denotes the kinetic-energy operator plus the external potential,
$\mu$ is the chemical potential,
$\beta\!\equiv\! T^{-1}$,
$\hat{1}$ is the unit matrix, and $\hat{H}_{0}$ and $\hat{1}(\omega_{n})$ are
defined by
\begin{equation}
\hat{H}_{0}
=\left[
\begin{array}{cc}
H_{0} & 0
\\
0 & -H_{0}^{*}
\end{array}
\right], \hspace{2mm}
\hat{1}(\omega_{n})
=\left[
\begin{array}{cc}
{\rm e}^{i\omega_{n}0_{+}} & 0
\\
0 & {\rm e}^{-i\omega_{n}0_{+}}
\end{array}
\right] ,
\end{equation}
with $0_{+}$ an infinitesimal positive constant.
The symbol
Tr in eq.\ (\ref{LWF}) also includes an integration over space variables
with multiplications of $\hat{H}_{0}$, $\hat{\tau}_{3}$, and $\hat{1}$
by the unit matrix $\delta({\mib r}\!-\!{\mib r}')$ implied.
Finally, $\hat{\Sigma}$ is the irreducible self-energy
matrix
obtained from the functional $\Phi\!=\!\Phi(\Psi,\Psi^{*},\hat{G})$
by
\begin{equation}
\hat{\Sigma}({\mib r},{\mib r}',\omega_{n})
\!=\!-2\beta\,  {\delta \Phi}/{\delta \hat{G}({\mib r}',{\mib r},\omega_{n})} \, .
\label{SigmaDef}
\end{equation}
With this relation, $\Omega$ is stationary with respect to
variation in $\hat{G}$ satisfying Dyson's equation:
\begin{equation}
\hat{G}^{-1}= i\omega_{n}\hat{1}-\hat{H}_{0}-\hat{\Sigma}+\mu\hat{\tau}_{3} \, .
\label{Dyson}
\end{equation}
A key quantity in eq.\ (\ref{LWF}) is $\Phi$.
We choose it so that the conservation laws and 
the Hugenholtz-Pines theorem are simultaneously satisfied.
Explicitly, it is given by
\begin{eqnarray}
&&\hspace{-11mm}
\Phi =\int\! d{\mib r}\! \int\! d{\mib r}'\,
U({\mib r}\!-\!{\mib r}')
\left\{ \frac{1}{2}\, |\Psi({\mib r})|^{2}|\Psi({\mib r}')|^{2}
\right.
\nonumber \\
&& \hspace{-5mm}
-\frac{1}{\beta} \! \sum_{n} \left[
|\Psi({\mib r})|^{2}\,
\frac{1}{2}{\rm Tr}\, \hat{\tau}_{3}\hat{G}({\mib r}',{\mib r}',\omega_{n})
\hat{1}(\omega_{n})\right.
\nonumber \\
&& \hspace{0mm} +\left.
\frac{1}{2}{\rm Tr}\, \hat{\tau}_{3} 
\mbox{\boldmath $\Psi$}({\mib r})
\mbox{\boldmath $\Psi$}^{\dagger} ({\mib r}')
\hat{G}({\mib r}',{\mib r},\omega_{n})
\hat{1}(\omega_{n}) \right]
\nonumber \\ 
&& \hspace{-5mm}
+ \frac{1}{2\beta^{2}} \! \sum_{n,n'} \left[
\frac{1}{2}{\rm Tr}\,\hat{\tau}_{3}\hat{G}({\mib r},{\mib r}, \omega_{n})
\hat{1}(\omega_{n})\right.
\nonumber \\
&& \hspace{11mm}\times
\frac{1}{2}{\rm Tr}\, \hat{\tau}_{3}\hat{G}({\mib r}',{\mib r}',\omega_{n'})
\hat{1}(\omega_{n'})
\nonumber \\
&& \hspace{0mm}
+ \left.\frac{1}{2}{\rm Tr}\, \hat{G}({\mib r},{\mib r}',\omega_{n})
\hat{1}(\omega_{n})
\hat{G}({\mib r}',{\mib r},\omega_{n'})
\hat{1}(\omega_{n'})\right] ,
\label{Phi}
\end{eqnarray}
where $U$ denotes the interaction potential and 
$\mbox{\boldmath $\Psi$}$ and 
$\mbox{\boldmath $\Psi$}^{\dagger}$
are defined in the same way as eq.\ (\ref{phi}).
The first and second terms in the two square brackets of eq.\ (\ref{Phi})
are the Hartree and Fock terms in Nambu space, respectively.
The difference between this theory and the HFB theory lies in the Fock terms. 
Indeed,
the HFB theory \cite{HM65,Griffin96} can be reproduced from eq.\ (\ref{Phi})
by replacing $\hat{G}$ and
$\hat{\tau}_{3} 
\mbox{\boldmath $\Psi$}({\mib r})
\mbox{\boldmath $\Psi$}^{\dagger} ({\mib r}')$ 
in the two Fock terms by
$\hat{\tau}_{3} \hat{G}$ and
$\mbox{\boldmath $\Psi$}({\mib r})
\mbox{\boldmath $\Psi$}^{\dagger} ({\mib r}')$, respectively;
the present functional was found by applying the reverse procedure.

Now, the self-energy can be calculated explicitly
with eq.\ (\ref{SigmaDef}), yielding the $(1,1)$ and $(1,2)$ 
components as
\begin{subequations}
\label{SD}
\begin{eqnarray}
&& \hspace{-13mm} \Sigma({\mib r},{\mib r}')
 =\delta({\mib r}\!-\!{\mib r}')
\int \!d{\mib r}''\,
U({\mib r}\!-\!{\mib r}'')\left[\,|\Psi({\mib r}'')|^{2}
\right.
\nonumber \\ 
&&\left.\hspace{35mm}+
\langle \phi^\dagger({\mib r}'')\phi({\mib r}'')\rangle\right]
\nonumber \\ 
&&\hspace{0mm}+ U({\mib r}\!-\!{\mib r}')\left[\Psi({\mib r})\Psi^*({\mib r}')+
\langle \phi^\dagger({\mib r}')\phi({\mib r})\rangle\right] ,
\label{Sigma}
\end{eqnarray}
\begin{equation}
\Delta({\mib r},{\mib r}')
 = U({\mib r}\!-\!{\mib r}')\left[\Psi({\mib r})\Psi({\mib r}')+
\langle \phi({\mib r})\phi({\mib r}')\rangle\right] ,
\label{Delta}
\end{equation}
\end{subequations}
respectively.
The $(2,1)$ and $(2,2)$ components are given by $-\Delta^{*}({\mib r},{\mib r}')$
and $-\Sigma^{*}({\mib r},{\mib r}')$, respectively.
We can also derive the equation for $\Psi({\mib r})$ 
from $\delta\Omega /\delta \Psi^{*}({\mib r})\!=\! 0$.
Noting $\delta\Omega/\delta\hat{G}\!=\!\hat{0}$,
we only need to consider
the explicit $\Psi^{*}$ dependences to obtain
\begin{equation}
(H_{0}\!-\!\mu)\Psi({\mib r})+\int\!\left[
\Sigma({\mib r},{\mib r}')\Psi({\mib r}')\!-\!\Delta({\mib r},{\mib r}')
\Psi^{*}({\mib r}')\right] d{\mib r}'= 0\, .
\label{GP}
\end{equation}
In the uniform case with no external potential where
$\Psi\!=\! {\rm constant}$, 
eq.\ (\ref{GP}) reduces to the
Hugenholtz-Pines relation 
$\mu\!=\! \Sigma_{{\mib k}={\mib 0}}\!-\!\Delta_{{\mib k}={\mib 0}}$,
as desired.
The expression for the particle number $N\!=\!-\partial\Omega/\partial \mu$ 
is obtained similarly as 
\begin{equation}
N=\int \! d{\mib r}\! \left[|\Psi({\mib r})|^{2}+
\langle \phi^\dagger({\mib r})\phi({\mib r})\rangle\right] .
\label{N}
\end{equation}
In order to calculate thermodynamic quantities, 
it is convenient to diagonalize the Green's function of eq.\ (\ref{Dyson}).
To this end, consider the following eigenvalue problem:
\begin{subequations}
\label{QP}
\begin{equation}
\int \!  \hat{H}({\mib r},{\mib r}')
\left[
\begin{array}{c}
u_{\nu}({\mib r}') \\ -v_{\nu}^{*}({\mib r}')
\end{array}
\right] d{\mib r}' = E_{\nu}
\left[
\begin{array}{c}
u_{\nu}({\mib r}) \\ -v_{\nu}^{*}({\mib r})
\end{array}
\right]  \, ,
\label{BdG}
\end{equation}
where $\hat{H}\!\equiv\!\hat{H}_{0}+ \hat{\Sigma}-\mu\hat{\tau}_{3}$, 
the subscript $\nu$ specifies the eigenstate,
and 
$(u_{\nu},v_{\nu})$ should be normalized as
\begin{equation}
\int \!\left[|u_{\nu}({\mib r})|^{2}\! -\! |v_{\nu}({\mib r})|^{2}\right] d{\mib r} =1 \, .
\label{Norm}
\end{equation}
\end{subequations}
One can show that $E_{\nu}$ is real when eq.\ (\ref{Norm}) 
can be satisfied \cite{Fetter72}.
It also follows from eq.\ (\ref{GP}) that, without the condition (\ref{Norm}),
eq.\ (\ref{BdG}) has a special solution $E \!=\! 0$
for $u \!=\! v \!=\!\Psi$.
We hence expect $E_{\nu}\!>\!0$ under the condition
(\ref{Norm}) if the system is stable;
the appearance of a negative eigenvalue evidences
an instability of the assumed $\Psi({\mib r})$.
Finally, one can show by using the symmetry 
$\hat{H}\!=\!-\hat{\tau}_{1}\hat{H}^{*}\hat{\tau}_{1}$
that the eigenstate of $\hat{H}$ corresponding to $-E_{\nu}$ 
is given by $(-v_{\nu},u_{\nu}^{*})^{\rm T}$.
Now, we can provide explicit expressions for
$\langle\phi^{\dagger}({\mib r}')\phi({\mib r})\rangle\!
=\!-{G}_{11}({\mib r},{\mib r}',\tau\!=\!-0_{+})$ and 
$\langle\phi({\mib r})\phi({\mib r}')\rangle\!
=\!-{G}_{12}({\mib r},{\mib r}',0)$ as
\begin{subequations}
\label{GF}
\begin{equation}
\langle\phi^{\dagger}({\mib r}')\phi({\mib r})\rangle\!=\!
\sum_{\nu}\left[u_{\nu}({\mib r})u_{\nu}^{*}({\mib r}')f_{\nu}\!+\!
v_{\nu}({\mib r})v_{\nu}^{*}({\mib r}')(1\!+\!f_{\nu})\right]  ,
\label{G}
\end{equation}
\begin{equation}
\langle\phi({\mib r})\phi({\mib r}')\rangle\!=\!
\sum_{\nu}\left[u_{\nu}({\mib r})v_{\nu}({\mib r}')\!+\!
v_{\nu}({\mib r})u_{\nu}({\mib r}')
\right](0.5\!+\!f_{\nu}) \, ,
\label{F}
\end{equation}
\end{subequations}
where $f_{\nu}\!\equiv
\!({\rm e}^{E_{\nu}/T}\!-\! 1)^{-1}$
is the Bose distribution function.
In deriving eq.\ (\ref{F}), use has been made of the identity
$\sum_{\nu}u_{\nu}({\mib r})v_{\nu}({\mib r}')\!=\!
\sum_{\nu}v_{\nu}({\mib r})u_{\nu}({\mib r}')$,
as can be proved from the completeness of the eigenfunctions
from eq.\ (\ref{QP}).

Equations (\ref{SD})-(\ref{GF}) form a closed 
set of self-consistent equations satisfying
both the Hugenholtz-Pines theorem
and various conservation laws.
Also, the pair correlation $\langle\phi\phi\rangle$
is adequately included in eqs.\ (\ref{GP}) and
(\ref{QP});
neglecting this contribution
yields the HFB-Popov theory.

Using eqs.\ (\ref{SD})-(\ref{GF}) and
the procedure in $\S 6$ of ref.\ \citen{Kita96},
it is possible to transform eq.\ (\ref{LWF}) 
in equilibrium 
into an expression without
Green's functions as
\begin{eqnarray}
&&\hspace{-7mm}
\Omega_{\rm eq} = T \sum_{\nu} 
\ln(1\!-\! {\rm e}^{-E_{\nu}/T})-  \sum_{\nu} 
E_{\nu}\int\! |v_{\nu}({\mib r})|^{2}\, d{\mib r}
\nonumber \\
&& \hspace{-3mm}
- \frac{1}{2}\int \! d{\mib r}\!\int \! d{\mib r}'\biggl\{
\Sigma({\mib r},{\mib r}')
\left[\Psi({\mib r}')\Psi^*({\mib r})+
\langle \phi^\dagger({\mib r})\phi({\mib r}')\rangle\right]
\nonumber \\
&& \hspace{9mm}
-\Delta^{\!*}({\mib r},{\mib r}')
\left[\Psi({\mib r}')\Psi({\mib r})+
\langle \phi({\mib r}')\phi({\mib r})\rangle\right]\biggr\}
\, .
\label{Omega}
\end{eqnarray}
The expression for the entropy can be obtained from 
$S\!=\!-\partial \Omega/\partial T$
by differentiating eq.\ (\ref{LWF})
in terms of the explicit 
$T$ dependences.
It yields the well-known expression:
\begin{equation}
S=\sum_{\nu}\left[(1+f_{\nu})\ln (1+f_{\nu})-f_{\nu}\ln f_{\nu}\right]\, .
\label{S}
\end{equation}
This completes the formulation of our mean-field theory.

Now, let us apply eqs.\ (\ref{SD})-(\ref{GF}) to a uniform system
with no external potential and the interaction:
\begin{equation}
U({\mib r}\!-\!{\mib r}')\!=\! 
\frac{4\pi a}{m}
\delta({\mib r}\!-\!{\mib r}') \, ,
\label{V}
\end{equation}
where $m$ is the particle mass
and $a$ is the $s$-wave scattering length.
We first expand quantities
dependent on a single argument ${\mib r}$
and two arguments $({\mib r},{\mib r}')$ 
in
$V^{-1/2}{\rm e}^{i{\mib k}\cdot{\mib r}}$ and
$V^{-1}{\rm e}^{i{\mib k}\cdot({\mib r}-{\mib r}')}$,
respectively,
with $V$ the volume.
We then find from eq.\ (\ref{SD}) that the expansion coefficients
$\Sigma_{\mib k}$ and $\Delta_{\mib k}$ are independent of ${\mib k}$;
we hence put 
$\Sigma_{\mib k}\!\rightarrow\!\Sigma$ and $\Delta_{\mib k}\!\rightarrow\!\Delta$
hereafter.
Also, $\Psi\!=\! \sqrt{n_{0}}$ 
($n_{0}$: condensate density) in eq.\ (\ref{GP}) 
so that the equation reduces
to the Hugenholtz-Pines relation $\mu\!=\!\Sigma\!-\!\Delta$.
Substituting these terms into eq.\ (\ref{QP}), 
we obtain
$E_{\mib k}\!=\!\sqrt{\epsilon_{\mib k}(\epsilon_{\mib k}\!+\! 2\Delta)}$,
$u_{\mib k}\!=\!\sqrt{(\xi_{\mib k}\!+\! E_{\mib k})/2E_{\mib k}}$, and
$v_{\mib k}\!=\!\sqrt{(\xi_{\mib k}\!-\! E_{\mib k})/2E_{\mib k}}$
with $\epsilon_{\mib k}\!=\! k^{2}/2m$ and
$\xi_{\mib k}\!=\!\epsilon_{\mib k}\!+\!\Delta$.
Putting these terms back into eqs.\ (\ref{SD}) and (\ref{N}),
we obtain $\Sigma\!=\! 8\pi n a/m$ with
$n\!\equiv\! N/V$ and
\begin{subequations}
\label{DN}
\begin{equation}
n = n_{0} + \frac{1}{V}\sum_{\mib k} \left(
\frac{\xi_{\mib k}\!-\! E_{\mib k}}{2E_{\mib k}}
+\frac{\xi_{\mib k}}{E_{\mib k}}\,
\frac{1}{{\rm e}^{E_{\mib k}/T}-1}\right) ,
\label{DN0}
\end{equation}
\begin{equation}
\Delta = \frac{4\pi a}{m} \left[n_{0}
+\frac{1}{V}\sum_{\mib k}\frac{\Delta}{2E_{\mib k}}
\left(1+\frac{2}{{\rm e}^{E_{\mib k}/T}-1}\right)
\right] .
\label{DN1}
\end{equation}
\end{subequations}
The first term in the round brackets of eq.\ (\ref{DN1})
produces an ultraviolet divergence.
In accordance with eq.\ (\ref{V}) where
the low-energy scattering length $a$ is used \cite{Pethick02},
we now introduce an energy
cutoff $\epsilon_{c}$ to the integral in such a way that 
$T_{0}\!\ll\!\epsilon_{c}\!\ll\! 1/ma^{2}$.
Substituting $T_{0}\!=\! 3.31n^{2/3}/m$,
the condition is transformed into
$1\!\ll\! \epsilon_{c}/T_{0}\!\ll\! 0.3(an^{1/3})^{-2}$,
which can be well satisfied for $an^{1/3}\!\ll\! 1$.
Other summations in eq.\ (\ref{DN}) are convergent,
and eq.\ (\ref{DN}) is now simplified into
\begin{subequations}
\label{ND}
\begin{equation}
\frac{n_{0}}{n}=1-\left(\!\frac{T}{T_{0}}\!\right)^{\!\! 3/2}
g_{1}({\Delta}/{T}) \, ,
\label{ND0}
\end{equation}
\begin{eqnarray}
&&\hspace{-15mm}
\left(1-\frac{\sqrt{8ma^{2}\epsilon_{c}}}{\pi} \right)
\frac{\Delta}{T_{0}}
\nonumber \\
&& \hspace{-15mm}=2\zeta(3/2)^{\! 2/3}an^{1/3}
\left[\frac{n_{0}}{n}+\frac{\sqrt{2\pi\Delta}\,T}{\zeta(3/2)T_{0}^{3/2}}\,
g_{2}({\Delta}/{T})\right]  ,
\label{ND1}
\end{eqnarray}
\end{subequations}
where $\zeta(3/2)$ is the Riemann $\zeta$ function, and
\begin{subequations}
\label{g12}
\begin{eqnarray}
&&\hspace{-5mm}g_{1}(x)\equiv\frac{2\sqrt{2}\,x^{3/2}}{\sqrt{\pi}\,\zeta(3/2)}\left(
2\!\int_{0}^{\infty}\!
\frac{\cosh 2t \sinh t}{{\rm e}^{x\sinh 2t}-1}\,dt +\frac{1}{3}\right)
\nonumber \\
&& \hspace{4mm}
\rightarrow 1-\frac{\sqrt{2\pi}}{\zeta(3/2)} \sqrt{x} +O(x) 
\hspace{5mm} (x\rightarrow 0) 
\, ,
\label{g1}
\end{eqnarray}
\begin{eqnarray}
&&\hspace{-6mm}g_{2}(x)\equiv\frac{4}{\pi}\,x\!\int_{0}^{\infty}\!
\frac{\sinh t}{{\rm e}^{x\sinh 2t}-1}\,dt 
\nonumber \\
&& \hspace{3mm}
\rightarrow 1-1.165\sqrt{x} +O(x) 
\hspace{5mm} (x\rightarrow 0) 
\, .
\label{g2}
\end{eqnarray}
\end{subequations}
Note that both $g_{1}$ and $g_{2}$
are given in powers of $\sqrt{x}$ for $x\!\rightarrow\! 0$,
which causes anomalous behavior near $T_{c}$
in various thermodynamic quantities, as seen below.
Equation (\ref{ND}) enables the complete
determination of $n_{0}$ and $\Delta$, thereby
fixing the thermodynamic equilibrium at a given temperature.
The chemical potential is then obtained by 
the Hugenholtz-Pines relation $\mu\!=\! 8\pi n a/m\!-\!\Delta$.
Other quantities of interest are the entropy $S$ of eq.\ (\ref{S})
and the superfluid density $\rho_{\rm s}$
obtained by \cite{Fetter72}
\begin{equation}
\rho_{\rm s}=n\left[1-\frac{1}{3mTV}\sum_{{\mib k}}
k^{2}\frac{{\rm e}^{E_{\mib k}/T}}{({\rm e}^{E_{\mib k}/T}-1)^{2}}\right] .
\label{rhoS}
\end{equation}

The transition temperature corresponds to the point where $n_{0}\!=\! 0$
in eq.\ (\ref{ND}).
Solving the resulting coupled equations for $T_{c}$ and 
$\Delta(T_{c})$ to the leading order in $\delta\!\equiv\! an^{1/3}$, 
we obtain
\begin{subequations}
\begin{equation}
\frac{T_c}{T_0}=1+\frac{8\pi}{3\zeta(3/2)^{4/3}}\delta=1+2.33\delta \, ,
\label{Tc}
\end{equation}
\begin{equation}
\frac{\Delta(T_{c})}{T_0}=\frac{8\pi}{\zeta(3/2)^{2/3}}\delta^2= 13.3\delta^2\, .
\label{Dc}
\end{equation}
Thus, the transition is first order with $\Delta(T_{c})\!>\!0$; it
is caused by the change in the low-energy quasiparticle
dispersion through $T_{c}$.
In contrast, $n_{0}$
is continuous at $T_{c}$ in the present theory.
Note the differences from the HFB-Popov theory
where $T_{c}\!=\!T_{0}$ and $n_{0}$ is discontinuous at $T_{c}$.
Our expression for $\Delta T_{c}$ agrees with 
the analytic result by Baym {\em et al}.\ \cite{Baym00}
as well as the numerical one by 
Holzmann and Krauth \cite{Holzmann99}.
Quantities other than $n_{0}$ are discontinuous
at $T_{c}$ as
\begin{equation}
\frac{\Delta\mu(T_{c})}{T_{0}}
=\frac{2}{3}\frac{\Delta S(T_{c})}{N}=
\frac{4\pi}{\zeta(3/2)^{2/3}}\delta^{2}  =6.63 \delta^{2}\, ,
\end{equation}
\begin{equation}
\frac{\Delta\rho_{s}(T_{c})}{nm}=\frac{4\pi}{3\zeta(3/2)^{4/3}}\delta
=1.16 \delta \, .
\end{equation}
\end{subequations}
Singular behaviors are also seen around $T\!\lesssim\! T_{c}$
stemming from $g_{1}$ and $g_{2}$ in eq.\ (\ref{g12}).
For example,
\begin{subequations}
\begin{equation}
\frac{n_{0}}{n} = \frac{\sqrt{6\pi}}{\zeta(3/2)^{2/3}} \delta^{1/2}(1-T/T_c)^{1/2}\, ,
\label{n0TTc}
\end{equation}
\begin{equation}
\frac{\Delta}{T_0} = \frac{\Delta(T_{c})}{T_0} +\sqrt{96\pi} \delta^{3/2}
(1-T/T_c)^{1/2} \, ,
\label{DTTc}
\end{equation}
\begin{equation}
\frac{S}{N} = \left(\! \frac{T}{T_{0}}\!\right)^{\! 3/2}
\left[\, \frac{5\zeta(5/2)}{2\zeta(3/2)}-\frac{3\Delta}{2T}\,\right] .
\label{STTc}
\end{equation}
\end{subequations}
It follows from eqs.\ (\ref{DTTc}) and (\ref{STTc}) that
the specific heat $C=T(\partial S/\partial T)$ just below
$T_{c}$ is divergent as
$C/N\! \sim \!13.0 \delta^{3/2}(1\!-\! T/T_c)^{-1/2}$.

Similar calculations at $T\!=\!0$ lead to the following expressions
in agreement with the Bogoliubov theory \cite{Fetter72}:
\begin{subequations}
\begin{equation}
\frac{n_{0}(0)}{n} = 1-\frac{8}{3\sqrt{\pi}} \delta^{3/2}
= 1-1.50 \delta^{3/2}\, ,
\end{equation}
\begin{equation}
\frac{\mu(0)}{T_{0}} =\frac{\Delta(0)}{T_{0}} 
= 2\zeta(3/2)^{2/3} \delta= 3.79 \delta\, .
\end{equation}
\end{subequations}
\begin{figure}[t]
\begin{center}
\includegraphics[width=6cm]{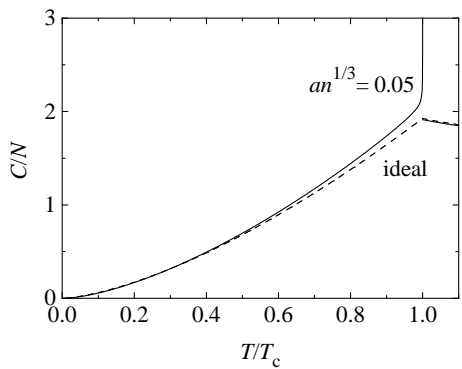}%
\end{center}
\caption{Specific heat $C/N$ as a function of $T/T_c$
for $an^{1/3}\!=\! 0.05$ shown in comparison with the ideal gas result.
}
\label{fig:1}
\end{figure}

\begin{figure}[t]
\begin{center}
\includegraphics[width=6cm]{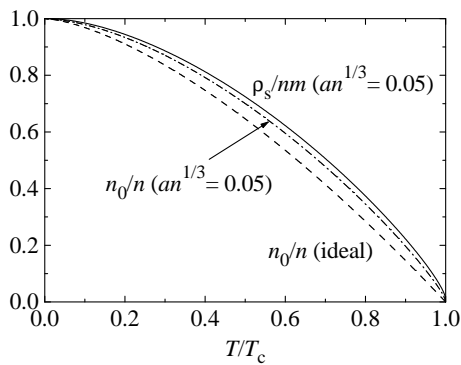}%
\end{center}
\caption{Normalized condensate density $n_{0}/n$
and superfluid density $\rho_{\rm s}/nm$ as functions
of $T/T_c$ for $an^{1/3}\!=\! 0.05$ shown
in comparison with the ideal gas result for 
$n_{0}/n\,(=\!\rho_{\rm s}/nm)$.
}
\label{fig:2}
\end{figure}

\begin{figure}[t]
\begin{center}
\includegraphics[width=6.5cm]{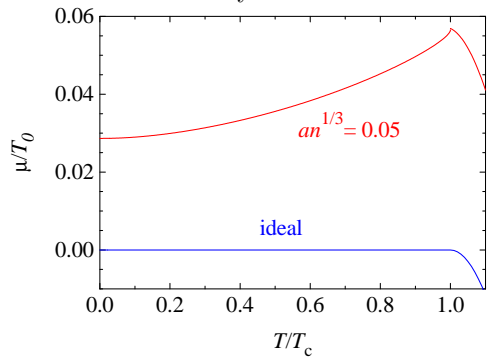}%
\end{center}
\caption{Temperature dependence of the chemical potential $\mu$
for $an^{1/3}\!=\! 0.05$ shown in comparison with the ideal gas result.
}
\label{fig:3}
\end{figure}

Figures \ref{fig:1}-\ref{fig:3}
show the overall temperature
dependences of $C/N$, $n_0/n$, $\rho_{\rm s}/nm$, and $\mu/T_{0}$
as functions of $T/T_c$ for a weakly interacting Bose gas 
with $an^{1/3}\!=\! 0.05$.
For comparison, the corresponding results 
for the ideal gas ($a\!=\! 0; T_c\!=\! T_0$)
are also plotted.
The curves for $an^{1/3}\!=\! 0.05$
satisfy the limiting behaviors 
derived above for $T\!\rightarrow\! T_c$ and $0$, 
although $\Delta\rho_{\rm s}(T_{c})$ and 
$1\!-\!n_{0}(0)/n$ are too small on the 
present scale to be seen clearly.

An important feature of the present mean-field theory is that
the thermodynamic equilibrium is fixed by the coupled eq.\
(\ref{ND}) for $n_{0}$ and $\Delta$.
Thus, the pair correlation $\langle \phi \phi\rangle
$ also plays an essential role
for $T_{c}$.
Now, one may ask: Are the results for the transition
compatible with the second-order
transition in superfluid $^4$He with $T_c\!<\!T_0$?
It is expected that, as the interaction between particles is increased
from zero,
$T_{c}$ initially increases towards a maximum,
decreasing eventually below $T_0$, as rationalized by Gr\"uter
{\em et al}.\cite{Gruter97}.
Also, the transition may change its character 
during the course from first- to second-order.

We finally comment on our result $\Delta T_{c}\!=\!2.33an^{1/3}T_0$
in connection with preceding numerical estimates. 
Among them, the path-integral approach by Holzmann and Krauth \cite{Holzmann99}
seems the most reliable, since they calculated $\Delta T_{c}$
in terms of a correlation function in the noninteracting Bose gas
derived by an expansion
from $a\!=\! 0$.
They thereby obtained the result $\Delta T_{c}\!\sim\!2.3an^{1/3}T_0$
compatible with our analytical one.
On the other hand, a different value $\Delta T_{c}\!\sim\!1.3an^{1/3}T_0$ 
has been obtained from Monte Carlo calculations by Arnold and Moore \cite{Arnold01}
and by Kashurnikov {\em et al}.\ \cite{Kashurnikov01}.
However, they both adopted as a starting point an effective action with
a finite interaction, where the effects from the pair correlation
$\langle\phi\phi\rangle$ may not have been included appropriately.
Finally, it is worth mentioning that the higher-order corrections to eq.\ (\ref{Tc})
depend on the cut-off $\epsilon_{\rm c}$ and are hence model dependent.

In summary, we have derived new mean-field equations for BEC
as eqs.\ (\ref{SD})-(\ref{GF}),
which are applicable to trapped atomic gases at finite temperatures.
They have been applied to a uniform system 
to reveal its basic thermodynamic features,
particularly around $T_c$.

The author is grateful to N. Schopohl for informative conversations.
This work is supported in part by a Grant-in-Aid for Scientific Research 
from the Ministry of Education, Culture, Sports, Science and Technology
of Japan.


\end{document}